\documentclass{emulateapj}
\usepackage{xspace}
\usepackage{amsmath}
\usepackage{hyperref}
\usepackage{framed} 
\usepackage{txfonts}
\usepackage{epstopdf} 
\def\nmin{n_{\rm min}}
\def\rmin{\rho_{\rm min}}
\def\tsf{\tau_{\rm SF}}
\def\Msun{{\rm M}_\odot}
\def\scl#1{#1\,{\rm pc}}

\def\rhos#1{\overline{\langle{\dot{\rho}_*}\rangle}_{#1}}
\def\evisfrd{EVISFRD}

\def\dim#1{\mbox{\,#1}}

\def\hide#1{}

\begin{document}

\title{Emergence of the Kennicutt-Schmidt Relation from the Small-Scale SFR-Density Relation}

\author{Nickolay Y.\ Gnedin\altaffilmark{1,2,3}, Elizabeth J.\ Tasker\altaffilmark{4}, and Yusuke Fujimoto\altaffilmark{4}}
\altaffiltext{1}{Fermi National Accelerator Laboratory, Batavia, IL 60510, USA; gnedin@fnal.gov}
\altaffiltext{2}{Kavli Institute for Cosmological Physics, The University of Chicago, Chicago, IL 60637 USA;}
\altaffiltext{3}{Department of Astronomy \& Astrophysics, The
  University of Chicago, Chicago, IL 60637 USA} 
\altaffiltext{4}{Department of Physics, Faculty of Science, Hokkaido University, Kita-ku, Sapporo 060-0810, Japan}

\begin{abstract}
We use simulations of isolated galaxies with a few parsec resolution to explore the connection between the small-scale star formation rate -  gas density relation and the induced large-scale correlation between the star formation rate surface density and the surface density of the molecular gas (the Kennicutt-Schmidt relation). We find that, in the simulations, a power-law small-scale ``star formation law'' directly translates into an identical power-law Kennicutt-Schmidt relation. If this conclusion holds in the reality as well, it implies that the observed approximately linear Kennicutt-Schmidt relation must reflect the approximately linear small-scale ``star formation law''.
\end{abstract}


\section{Introduction}
\label{sec:intro}

By now it is reasonably well established that on large scales (many hundreds of parsecs) the star formation rate surface density correlates strongly and approximately linearly with the surface density 
of the CO-emitting gas (and, under the assumption of the constant $X_{\rm CO}$ factor, with the molecular gas; an area under scrutiny but beyond the scope of this paper). 
This relation, commonly called Kennicutt-Schmidt, holds both in the local universe \citep{ism:lwbb08,sfr:blwb08,sfr:bljo11,sfr:blwb11,sfr:lbbb12,sfr:lwss13} and at intermediate cosmic redshifts \citep{sfr:gtgs10,sfr:dbwd10,sfr:tngc13}. 

The origin of the large-scale linear Kennicutt-Schmidt relation is still not fully understood. Stars form on small scales, hence the large-scale behavior is the result of averaging of the intrinsic small-scale star formation ``law''. Such averaging can proceed in two distinct regimes: the large-scale linear behavior may be ``emergent'', i.e. irrespectively how star formation proceeds on small scales, large-scale averaging results in a linear law. For example, in cosmology it is well-known that on sufficiently large scales the galaxy bias is linear even if on small scales galaxy formation is a highly nonlinear process. If the same property holds for the density distribution in the ISM, it would provide a simple and elegant explanation to why the Kennicutt-Schmidt relation is (nearly) linear on large scales.

Alternatively, the ``memory'' of small-scale conditions can be preserved, in which case the linearity of the large-scale Kennicutt-Schmidt relation would imply the linear relation between the rate of star formation and gas density on small scales. 

The relationship between star formation rate (SFR) and gas density on small scales is understood even less. For example, \citet{sfr:lla10} found that on the scale of individual star-forming cores ($\la 1\dim{pc}$) the SFR - density relation is also linear, but only if the density is above the threshold of $700\Msun/\dim{pc}^{3}$. On the other hand, \citet{sfr:kt07} argued that a ``constant efficiency per free-fall time'' ($\dot{\rho}_*\propto\rho^{3/2}$) is, instead, a better interpretation of several observational measurements.

This apparent inconsistency could arise from mixing diverse spatial scales, so it is worth reviewing how the hierarchy of spatial scales affects the SFR - density relation.

Throughout this paper we try to be careful with our terminology. We reserve the term ``Kennicutt-Schmidt relation'' for the relationship between the surface densities of stars and (molecular) gas, and only on large scales (since surface densities are only meaningfully defined on scales much larger than the disk scale height). When we compare the volumetric star formation rate density with gas density, we use the term ``SFR - density relation''.

\section{SFR - density relation on various scales}
\label{sec:avg}

It is useful to think about relationship between star formation rate and gas densities (on any scale) in terms of the time-scale for gas consumption, the so-called ``depletion time'' $\tsf$,
\[
  \dot{\rho}_* = \frac{\rho}{\tsf}.
\]
The above equation is, however, an incorrect way of considering star formation, because the density is \emph{only defined} on a particular scale - we need to explicitly consider the range of spatial scales that is relevant for our problem.

Let us take some spatial scale $L$. One can imagine a region of interest (be it the whole universe or a single galaxy) divided into boxes of size $L$. If we average gas densities on scale $L$, they become meaningfully defined. Thus, the above equation could be replaced with
\[
  \langle\dot{\rho}_*\rangle_L = \frac{\langle\rho\rangle_L}{\tsf},
\]
but even that is not general enough. Star formation rate is not a deterministic function of gas density, but dependent on other physical parameters (previous history, dynamic state of the gas, etc); it needs to be treated as a stochastic process, described by a probability distribution of having a particular value of the instantaneous star formation rate density for a given gas density at scale $L$, $p\left(\langle\dot{\rho}_*\rangle_L|\langle\rho\rangle_L\right)$.

For any form of $p$, however, it is possible to define the average instantaneous star formation rate density at a given gas density,
\begin{equation}
  \rhos{L}\left(\langle\rho\rangle_L\right) \equiv \int x p\left(x|\langle\rho\rangle_L\right) dx.
  \label{eq:evisfrd}
\end{equation}
Notation here becomes complex, so we use angular brackets $\langle\rangle_L$ to indicate averaging over a given spatial scale (a single box of size $L$), and the over-bar to mark the average over all boxes with the same $\langle\rho\rangle_L$; to sharpen the terminology, we also call the latter average "the expectation value of the instantaneous star formation rate surface density on scale $L$" (\evisfrd). With this notation and terminology, we can finally relate the \evisfrd\ on spatial scale $L$ to the gas density on the same scale,
\begin{equation}
  \rhos{L} = \frac{\langle\rho\rangle_L}{\tsf},
  \label{eq:tsfL}
\end{equation}
 (or, equivalently, $\langle\overline{\Sigma_*}\rangle_L = \langle\Sigma\rangle_L/\tsf$ when considering surface densities). In that case depletion time becomes the function of other gas properties on scale $L$,
\[
  \tsf = \tsf(L,\langle\rho\rangle_L,...).
\]
In other words, we need to explicitly consider the star formation relation as (at least) a two-dimensional relation on the plane $(L,\langle\rho\rangle_L)$.

In this formalism the large-scale Kennicutt-Schmidt relation can be expressed as
\[
  \tsf = \left.(2\pm1)\dim{Gyr}\,\right|_{L\gg\scl{100}},
\]
with substantial \emph{intrinsic} (i.e.\ exceeding the formal observational error) scatter and the actual value for $\tsf$ noticeably different at $z=0$ \citep[$\tsf\approx2\dim{Gyr}$][]{ism:lwbb08,sfr:blwb08,sfr:bljo11,sfr:blwb11,sfr:lbbb12,sfr:lwss13} and high redshift \citep[$\tsf\approx0.7\dim{Gyr}$][]{sfr:gtgs10,sfr:dbwd10,sfr:tngc13}. In this paper we are going to overlook these important details, since we are only interested in the broad-brush effect of varying spatial scale on the SFR -  density relation. Hence, for our purposes it is sufficient to consider $\tsf$ as constant on large scales.

\citet{sfr:lla10} observations can now be simply expressed as 
\[
  \rhos{\scl{1}} =
	\left\{
    \begin{array}{ll}
	  \displaystyle\frac{\langle\rho\rangle_{\scl{1}}}{20\dim{Myr}}, & \rho > \rmin, \\
	  0, & \rho < \rmin, \\
	\end{array}
	\right.
\]
with $\rmin=700\Msun/\dim{pc}^3$. For this equation to be consistent with the Kennicutt-Schmidt relation on large (say, $500\dim{pc}$) scale,
\[
  \rhos{\scl{500}} = \frac{\langle\rho\rangle_{\scl{500}}}{2\dim{Gyr}},
\]
exactly 1\% of the molecular gas must sit above the small-scale density threshold $\rmin$ - after all, 
\[
  \rhos{\scl{500}} = \left\langle\rhos{\scl{1}}\right\rangle_{\scl{500}}.
\]

In the language of depletion time the ``constant efficiency per free-fall time'' ansatz becomes
\[
  \tsf(L,\langle\rho\rangle_L,...) = 
  \frac{\tau_{\rm ff}(\langle\rho\rangle_L)}{\epsilon_{\rm SF}} = 
  \epsilon_{\rm SF}^{-1}\sqrt{\frac{3\pi}{32G\langle\rho\rangle_L}},
\]
or, in a more familiar form,
\begin{equation}
  \rhos{L} = \epsilon_{\rm SF}\frac{\langle\rho\rangle_L} 
  {\tau_{\rm ff}} = \epsilon_{\rm SF} \frac{\langle\rho\rangle_L^{3/2}}{\sqrt{3\pi/(32G)}}.
  \label{eq:tffsf}
\end{equation}
This ansatz is consistent with the observational constraints presented by \citet{sfr:kt07}. However, those constraints sample not only various densities, but also \emph{various spatial scales}. Indeed, the lowest densities $n_{\rm H}\ga 10\dim{cm}^{-3}$ in \citet{sfr:kt07} are sampled by CO measurements, that typically include whole molecular clouds on scales $L\sim 10-30\dim{pc}$; densities around $n_{\rm H}\sim 10^3\dim{cm}^{-3}$ are observed in Infrared Dark Clouds (IRDC), which have sizes $1-3\dim{pc}$; finally the highest densities of $n_{\rm H}\sim 10^5\dim{cm}^{-3}$ are only present in HCN-emitting star-forming cores with sizes around $0.1-0.3\dim{pc}$ or below. Hence, the observational constraints used by \citet{sfr:kt07} all fall along a particular track $L^2\times\langle\rho\rangle_L \sim (1-10)\times10^3\dim{cm}^{-3}\dim{pc}^2$ in the two-dimensional plane $(L,\langle\rho\rangle_L)$. In other words, observational constraints that support the ``constant efficiency per free-fall time'' ($\tsf\propto\langle\rho\rangle_L^{-1/2}$) equally well support the ``constant efficiency per unit scale'' ($\tsf\propto L$).

In fact, the following two models for the depletion time:
\begin{description}
\item{{\bf ``linear''}}
  \[
  \tsf(L,\langle\rho\rangle_L,...) \sim 
  \left\{
  \begin{array}{ll}
      2\dim{Gyr}\times\mbox{min}\left(1,\frac{L}{L_0}\right), & \langle\rho\rangle_L > \rmin, \\
      \infty, & \langle\rho\rangle_L < \rmin, \\
  \end{array}
  \right.
  \]
\item{{\bf ``3/2''}}
  \[
  \tsf(L,\langle\rho\rangle_L,...) \sim 
  \left\{
  \begin{array}{ll}
      100\tau_{\rm ff}(\langle\rho\rangle_L), & L < L_0, \\
      2\dim{Gyr}, & L > L_0, \\
  \end{array}
  \right.
  \]
\end{description}
with $\rmin \sim \rho_0\times\mbox{min}\left(1,L_0^2/L^2\right)$, cannot be distinguished at present without additional observational constraints. The parameter $L_0$ in both cases must be in the range of a few hundred parsecs (for example, the scale height of the gaseous disk) to be consistent with the large-scale Kennicutt-Schmidt relation. In the Milky Way galaxy $\rho_0$ is such that the \citet{sfr:lla10} result is matched ($\rho_{\rm min}(L)\approx700\Msun/\dim{pc}^{3}$ at $L\sim1\dim{pc}$), but in other galaxies it may be different (for example, being proportional to the density of the atomic-to-molecular transition).

\hide{
Figure \ref{fig:sf} shows the two alternatives in a cartoon fashion. At present, either one is a sensible model, as well as any other, intermediate or more complex model, that still matches the observational constraints.
}

Even if observations are not yet conclusive, we can at the very least explore the scale dependence of the SFR - density relation in high resolution numerical simulations. In fact, since we are only concerned with the effect of changing the spatial scale $L$ on the relation, we do not even need dynamic simulations - all that is required is a snapshot with a realistic gas distribution down to sufficiently small (few parsec) scales.

\section{Simulations}
\label{sec:sim}

\begin{table}[b]
\caption{Simulations\label{sim}}
\centering
\begin{tabular}{llll}
\hline\hline\\
Simulation & Resolution & SF/Feedback & Grand design \\
\hline\\
M83HighRes & 1.5 pc & no & yes \\
M83LowRes  & 6.0 pc & no & yes \\
MWLowRes   & 7.8 pc & yes & no \\
\\
\hline
\end{tabular}
\end{table}

In order to test such a conjecture, we use three adaptive mesh refinement simulations of isolated galaxies. For the first two simulations, the galaxy model is based on observations of M83 and has a grand design bar and spiral excited by the gravitational potential from particles which rotate at a fixed pattern speed. The simulations were run using the 3D hydrodynamical code, Enzo \citep{Enzo2013}, with radiative cooling down to 300\,K, and did not include active star formation or feedback. To prevent unresolved collapse, a pressure floor was introduced when the Jeans length covered less than 4 cell widths. The limiting resolution (smallest cell size) for the high resolution run, M83HighRes, was 1.5\,pc, and for the lower resolution realization, M83LowRes, it was 6\,pc. Further analysis of these simulations will be found in Fujimoto, Tasker, Wakayama, \& Habe 2014 (MNRAS, in press).

The third simulation is of a Milky Way-sized galaxy (MWLowRes) without an imposed grand design. The model initial conditions are presented in \citet{Tasker2011}; its spatial resolution is similar to the M83LowRes run, but the MWLowRes run includes star formation and supernovae feedback in the form of a thermal energy injection. 

A summary of the three simulations is given in Table 1.

\section{Results}
\label{sec:res}

\begin{figure}[t]
\includegraphics[width=1\hsize]{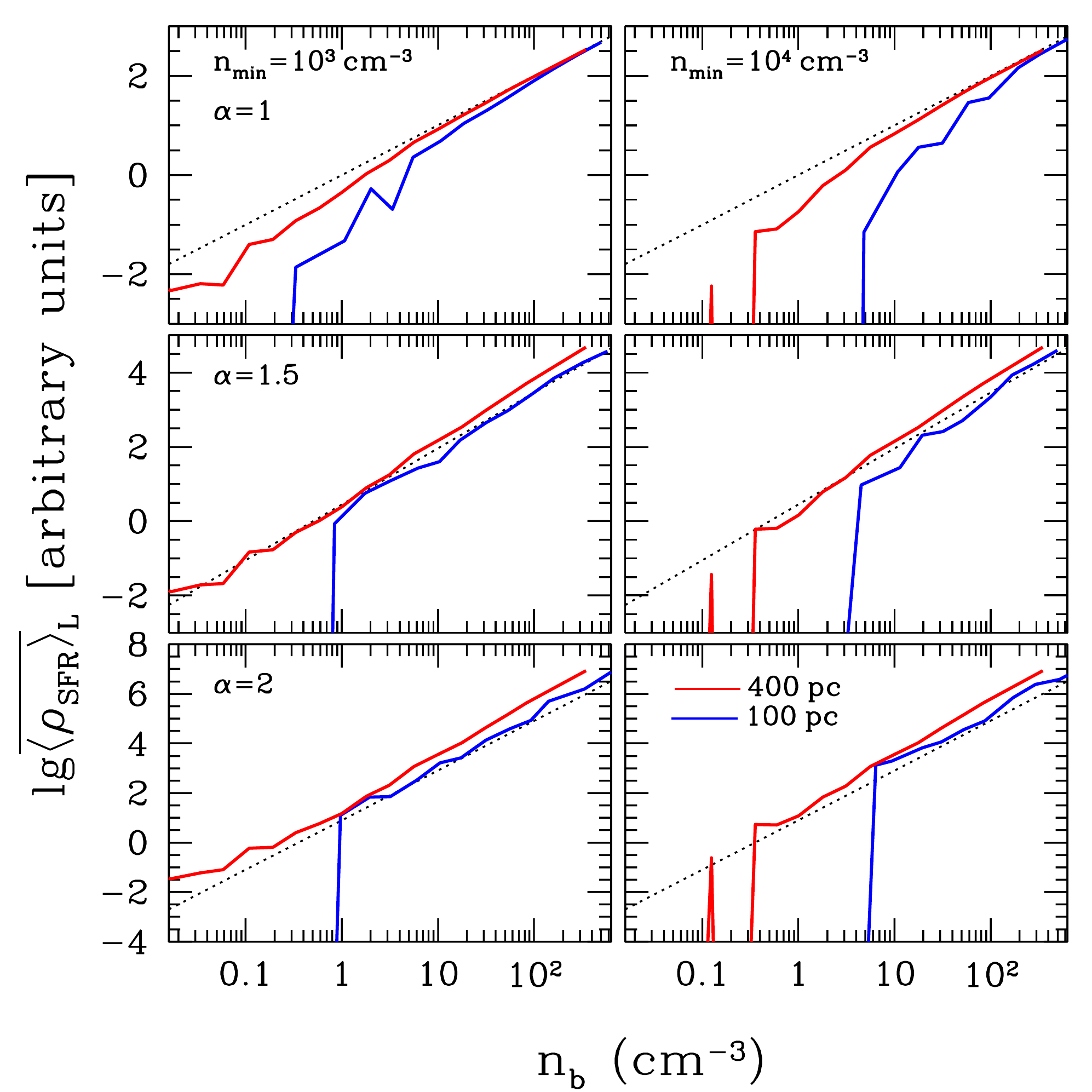}
\caption{Expectation value of the instantaneous star formation rate density (in arbitrary units) vs average gas density at two averaging scales ($100\dim{pc}$: blue, $400\dim{pc}$: red) for 6 star formation models with $n_{\rm min}=10^3,10^4\dim{cm}^{-3}$ and $\alpha=1,1.5,2$ for our fiducial simulation M83HighRes. In each row of panels black dotted line shows the respective power-law relation.}
\label{fig:den}
\end{figure}

In order to explore the effect of averaging in the simulations, we adopt a simple power-law-with-threshold relation between the \evisfrd\ $\rhos{l}$ and the gas density $\rho$ at the resolution scale of the simulation $l$. For the sake of brevity, we use just $\rho$ instead of $\langle\rho\rangle_l$ and $\dot{\rho}_*$ instead of $\rhos{l}$ for the density and the \evisfrd\ on the smallest resolved scale,
\begin{equation}
  \rhos{l} \equiv  \dot{\rho}_* \propto \begin{cases} \rho^\alpha, &\mbox{if } \rho \geq \rmin \\
0 & \mbox{if } \rho < \rmin. \end{cases}.
  \label{eq:rel}
\end{equation}
Since we are only interested in the relative changes in $\rhos{}$ with averaging over various scales, the normalization of Equation (\ref{eq:rel}) is not important to us, and we always measure the star formation rate in arbitrary units.

In the simulations, the molecular gas is not followed explicitly. However, the values for the density threshold we choose are always at or above $n_{\rm H}=10^3\dim{cm}^{-3}$, which, for even parsec scales, corresponds to molecular gas. At this value of the threshold the density PDF in the M83HighRes run already deviates strongly from lognormal, having developed the power-law tail induced by self-gravity, in agreement with turbulent ISM simulations \citep[e.g.][]{ism:knw11}. Hence, we only allow star formation in self-gravitating gas.

Starting with the simulation resolution scale $l$, we first assign the star formation rate $\dot{\rho}_*$ on the most refined cells throughout the whole computational domain. Then we simply average both $\rho$ and $\dot{\rho}_*$ independently over cells in groups of 8, building up the SFR - density relation sequentially on scales $L=2\times l$, then $L=4\times l$, etc. At each scale we compare $\rhos{L}$ versus $\langle\rho\rangle_L$; that relationship at two spatial scales is shown in Figure \ref{fig:den} for our fiducial simulation M83HighRes.

The immediately obvious conclusion is that the slope $\alpha$ of the small-scale power-law relation (\ref{eq:rel}) is preserved through the averaging procedure all the way to low densities. At the lowest densities the power-law scaling steepens, reflecting the density threshold, with the characteristic density until which the power-law scaling is maintained decreasing with increasing averaging scale. For sufficiently large averaging scales (many hundreds of parsecs), the power-law scaling extends well beyond threshold densities and even into the regime where the gas is expected to be mostly atomic. In other words, the density structure of the ISM (at least in the simulation) is \emph{not} like the density distribution of large-scale structure in the universe, on large scales the ``bias'' does \emph{not} become linear. 

Such a behavior likely indicates that the ISM density distribution is highly correlated in space. This is not particularly surprising, though, since in the real ISM density structures are, indeed, organized along spiral arms (either grand design or flocculent) and star forming sites are strongly clustered in their parent molecular clouds (notice, that the latter statement does not imply that the clustering properties of molecular clouds along spiral arms are similar to clustering properties of stars inside GMCs - they clearly cluster differently; all that matters is that they cluster strongly in each case, with their respective distributions, while being different from each other, also both being very different from a Gaussian random field).

\begin{figure}[t]
\includegraphics[width=1\hsize]{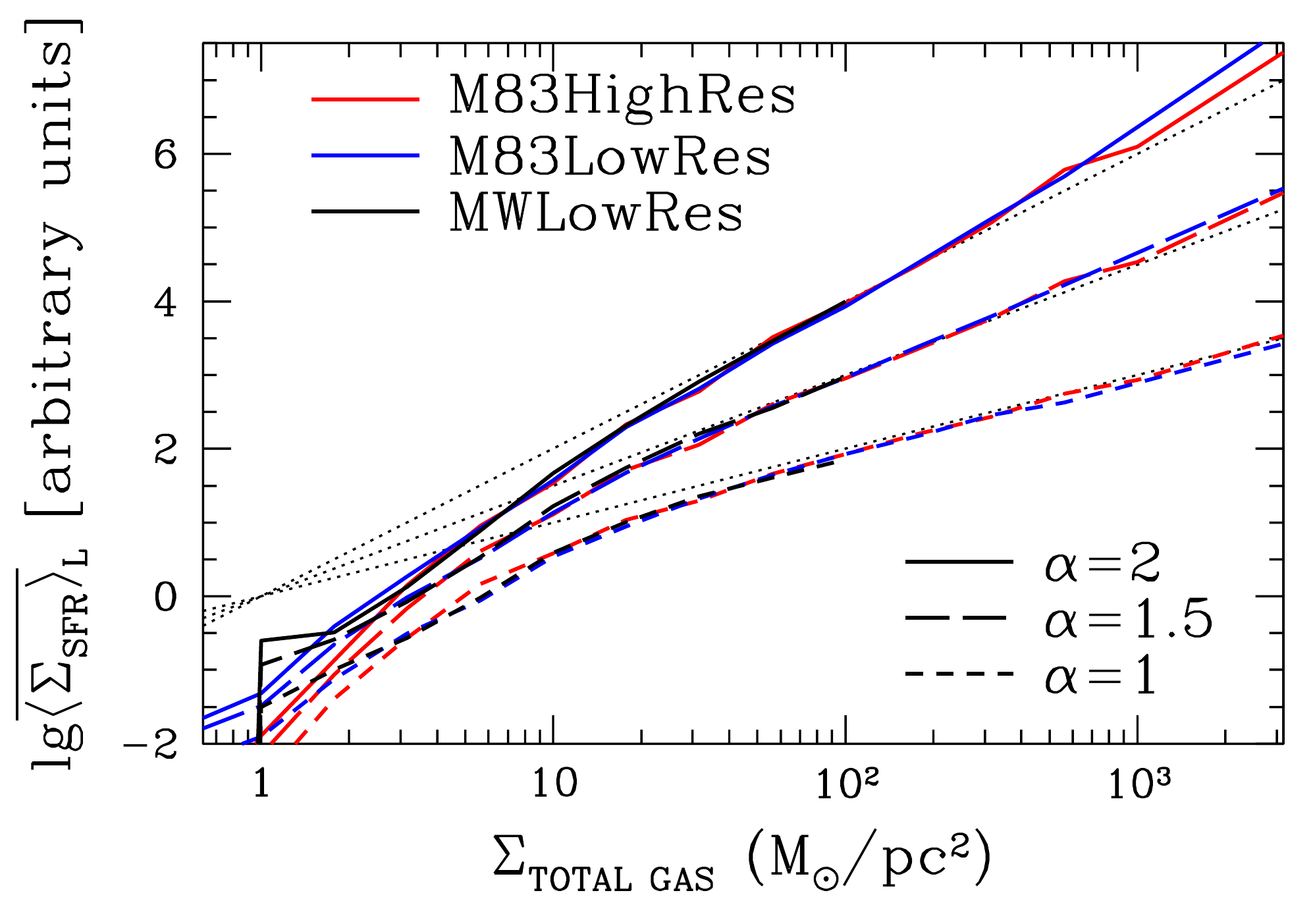}
\includegraphics[width=1\hsize]{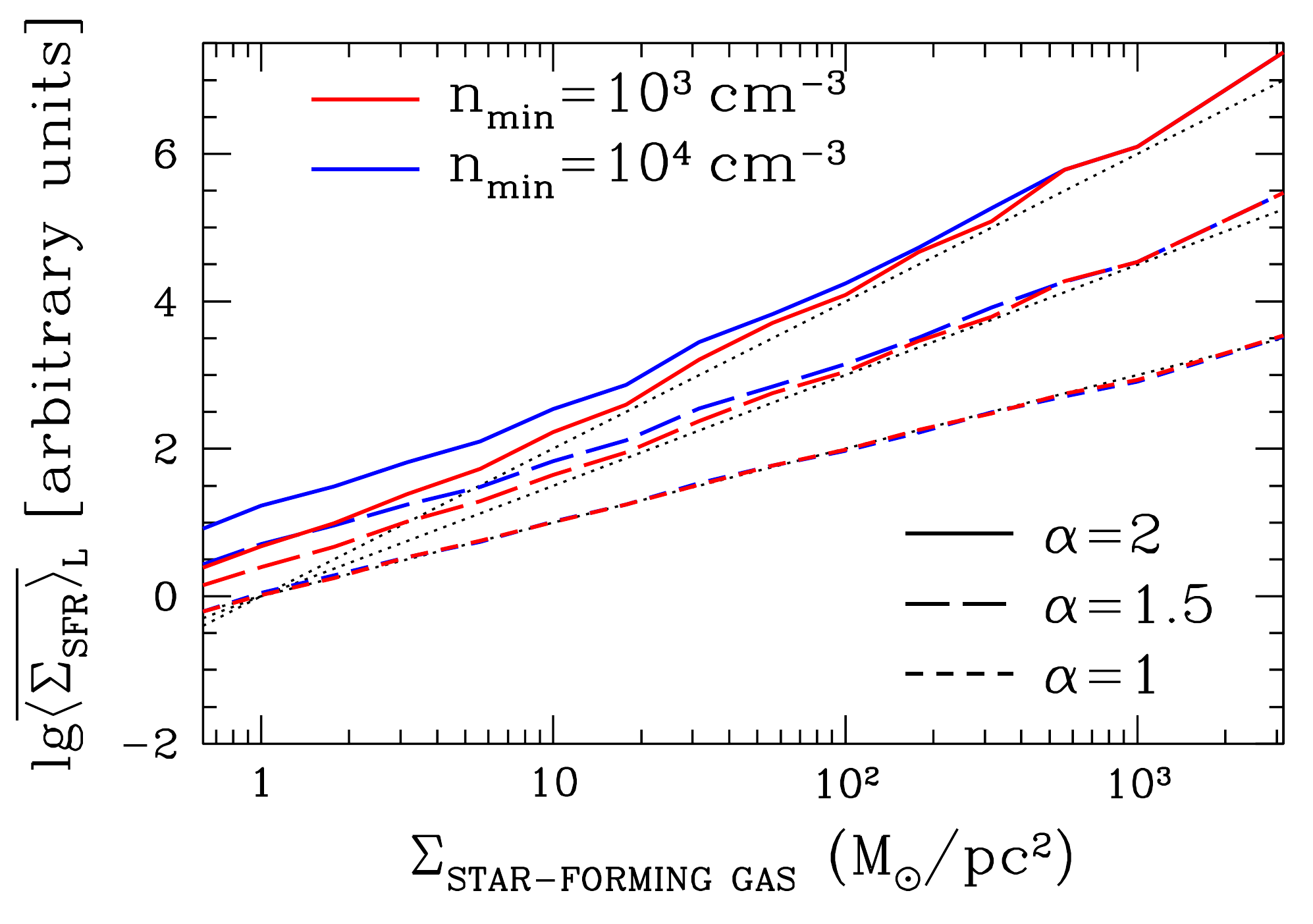}
\caption{Expectation value of the instantaneous star formation rate surface density (in arbitrary units) vs total gas (top) and star-forming gas (bottom) surface density at $400\dim{pc}$ averaging scale for three values of the small-scale slope of the star formation model $\alpha=1,1.5,2$. In the top panel red, blue, and black colors show three simulations : low resolution M83LowRes, high resolution M83HighRes (our fiducial one), and MWLowRes that includes star formation and supernova feedback; in the bottom panel the fiducial simulation M83HighRes is shown for two values of the density threshold:  $n_{\rm min}=10^3\dim{cm}^{-3}$ (red) and $n_{\rm min}=10^4\dim{cm}^{-3}$ (blue).}
\label{fig:sig}
\end{figure}

It is already clear from Figure\ \ref{fig:den} that the nonlinear small-scale relation should be preserved for surface densities as well (since at high densities/surface densities the relation is scale independent). Just to illustrate that, we show in the top panel of Figure \ref{fig:sig} the surface densities of gas and the \evisfrd\ on the largest averaging scale ($400\dim{pc}$) for $\nmin=10^3\dim{cm}^{-3}$ (it is apparent from  Figure \ref{fig:den} that large-scale behavior is weakly dependent on the density threshold $\nmin$) for three representative simulations. Indeed, the small-scale slope $\alpha$ is also preserved in the relationship between surface densities (at least until very low surface densities, well below observational detection thresholds).

While observations indicate the best correlation lies between star formation and CO emission, our model only follows atomic gas. We therefore can only use ‘'star-forming gas’' whose density is above our specified threshold, when plotting the KS relation. The steepening in the relation due to the threshold is, expectantly, absent in that case, but all the qualitative features of the relation are fully preserved in that case as well.

Figure\ \ref{fig:sig} also demonstrates the robustness of our results to changing numerical resolution and modeled physics: simulations M83HighRes and M83LowRes differ by a factor of 4 in spatial resolution, but are barely distinguishable above $\Sigma_{\rm gas}\approx5\Msun/\dim{pc}^2$; simulations M83LowRes and MWLowRes have comparable resolutions, but the first does not include the feedback, while the second one does, never-the-less, the results from them are fully consistent.

\section{Discussion}
\label{sec:cons}

Our main conclusion is that the large-scale Kennicutt-Schmidt relation is \emph{not} emergent, but, instead, directly inherits the functional form of the small-scale SFR - density relation. In the simulations this conclusion is robust - it holds for runs with and without feedback, and in simulations with varied resolution.

That conclusion can be formalized further if one considers explicitly the spatial averaging of equation (\ref{eq:rel}) as we move from the small scale $l$ to some large scale $L$. Spatial averaging inside each large box $L$ can be represented by an integral over a local density PDF,
\[
  \langle\rho\rangle_L(\vec{x}_i) = \int \rho\, {\rm PDF}(\rho,\vec{x}_i)\, dV,
\]
where $\vec{x}_i$ is the location of the center of a given box $i$, and the integral is taken over the volume $V=L^3$. Then for the \evisfrd\ $\rhos{L}$ we obtain (Eq.\ \ref{eq:evisfrd})
\begin{eqnarray}
   \rhos{L} & = & \frac{1}{N} \sum_i \int \dot{\rho}_*(\rho)\, {\rm PDF}(\rho,\vec{x}_i)\, dV\nonumber\\
                & = & \int \dot{\rho}_*(\rho)\, \overline{\rm PDF}(\rho|\langle\rho\rangle_L)\, dV,\nonumber
\end{eqnarray}
where $\overline{\rm PDF}(\rho|\langle\rho\rangle_L)$ is the average density PDF in our region of interest for all boxes with the mean density $\langle\rho\rangle_L$.

For our model ansatz (\ref{eq:rel}) we can, thus, obtain
\begin{eqnarray}
  \rhos{L} & \propto & \int_{\rmin}^\infty \rho^\alpha \overline{\rm PDF}(\rho|\langle\rho\rangle_L)\, dV\nonumber\\
               & = & \int_{0}^\infty \rho^\alpha \overline{\rm PDF}(\rho|\langle\rho\rangle_L)\, dV \times \left(1-f_{\rm low}\right),
  \label{eq:pdf}
\end{eqnarray}
where $f_{\rm low}$ is the relative contribution to the integral from densities below the threshold. We can now consider several special cases as illustrative examples. 

{\bf Example 1.} For $\alpha=1$, for any form of $\overline{\rm PDF}(\rho|\langle\rho\rangle_L)$, we have
\[
  \rhos{L} \propto \langle\rho\rangle_L \left(1-f_{\rm low}\right) \equiv  \langle\rho_{\rm SFG}\rangle_L,
\]
i.e.\ the small-scale linear SFR-density relation is preserved at large scales for the total gas Kennicutt-Schmidt relation in environments where $f_{\rm low}$ is small -  for example, when at progressively higher average densities $\langle\rho\rangle_L$ the PDF shifts towards progressively higher small-scale densities  (so that $f_{\rm low}$ decreases with increasing $\langle\rho\rangle_L$). At low enough densities, where $f_{\rm low}$ is not negligible, the Kennicutt-Schmidt relation becomes steeper than linear, as was first shown by \citet{misc:k03} and is also visible from Fig.\ \ref{fig:sig} \citep[see also][]{ng:fgk11,ng:fgk12b}. The Kennicutt-Schmidt relation for the star-forming gas then simply becomes mathematical identity (up to a  constant proportionality factor) in all environments and for any PDF (for example, that may explain the linear dependence of the star formation rate on HCN luminosity).

{\bf Example 2.} If the PDF is self-similar, $\overline{\rm PDF}(\rho|\langle\rho\rangle_L) = F(\rho/\langle\rho\rangle_L)$, as, for example, is often found in the simulations of the turbulent ISM \citep{sfr:mo07,ism:knw11,ism:pn11,ism:kln13},  then for any $\alpha$
\[
  \rhos{L} \propto \langle\rho\rangle_L^\alpha \left(1-f_{\rm low}\right),
\]
and, again the small-scale SFR-density relation directly propagates into the total gas Kennicutt-Schmidt relation in the limit $f_{\rm low} \ll 1$.

It is worth noting that since we consider only the instantaneous KS relation, we do not comment upon the effect of feedback processes that may alter the fraction of gas available for future star formation. Feedback has been proved to affect the gas distribution on time-scales of several tens Myr, causing self-regulation in the form of weak dependence of the Kennicutt-Schmidt relation on the star formation efficiency \citep{sims:svbw09,sims:atm11,sims:hqm11,ng:akl13,sims:hko13}, but this impacts the PDF from Eq. 6 which is independent of our results. Such a shift will affect the amplitude of the KS relation, but does not affect whether the slope is preserved over large scales.

In this paper we deliberately measured star formation in arbitrary units, to sidetrack any effect of varying efficiency and the whole question of the amplitude of the Kennicutt-Schmidt relation. Instead, we entirely focus on the \emph{slope} of the relation, which is indeed preserved on large scales in other simulations as well. For example, both \citet{ng:akl13} and \citet{sims:hko13} used a ``3/2'' model for the small-scale SFR - density relation, and found a similar slope of the Kennicutt-Schmidt relation on large scales, in full agreement with our results. 

Hence, irrespective of the strength or a specific implementation of the feedback model, in cosmological and galactic scale simulations the slope of the small-scale SFR - density relation is preserved in the large-scale Kennicutt-Schmidt relation.

Is this conclusion robust in real life? Barring unexpected problems with simulations, one should never forget that, while the observations measure the instantaneous value of the molecular gas density, any large-scale measure of star formation rate always returns a time averaged value - over a time scale of about $20\dim{Myr}$ for UV observations and about $4-5\dim{Myr}$ for H-$\alpha$ measurements. Hence, the observational determinations of the large-scale Kennicutt-Schmidt relation always compare apples and oranges, as was convincingly shown by, for example, \citet{sfr:slw10}. 

The mismatch between the measurements of the instantaneous  gas density and of the time-averaged star formation rate may, potentially, cause the emergence of the large-scale linear Kennicutt-Schmidt relation from a non-linear small-scale SFR-density relation (to the best of our knowledge, that has not been demonstrated, though, so we only mention this as a possibility). If that was the case, however, it would be really bad news for the observers, implying that all of the existing large-scale studies of the Kennicutt-Schmidt relation are, in some sense, misleading, as they obscure the true non-linear relation between the instantaneous gas density and the star formation rate.

Irrespectively of the proper interpretation of our results, it is clear that studying the linear small-scale relation between the SFR and molecular gas density is as important as exploring the ``constant efficiency per free-fall time'' ansatz.

\acknowledgements
We are grateful to Andrey Kravtsov, Diederik Kruijssen, and the anonymous referee for the constructive criticism that significantly improved the original manuscript.

\bibliographystyle{apj}
\bibliography{ms,ng-bibs/self,ng-bibs/ism,ng-bibs/misc,ng-bibs/sims,ng-bibs/sfr}

\begin{thebibliography}{26}
\expandafter\ifx\csname natexlab\endcsname\relax\def\natexlab#1{#1}\fi

\bibitem[{{Agertz} {et~al.}(2013){Agertz}, {Kravtsov}, {Leitner}, \&
  {Gnedin}}]{ng:akl13}
{Agertz}, O., {Kravtsov}, A.~V., {Leitner}, S.~N., \& {Gnedin}, N.~Y. 2013,
  \apj, 770, 25

\bibitem[{{Agertz} {et~al.}(2011){Agertz}, {Teyssier}, \& {Moore}}]{sims:atm11}
{Agertz}, O., {Teyssier}, R., \& {Moore}, B. 2011, \mnras, 410, 1391

\bibitem[{{Bigiel} {et~al.}(2008){Bigiel}, {Leroy}, {Walter}, {Brinks}, {de
  Blok}, {Madore}, \& {Thornley}}]{sfr:blwb08}
{Bigiel}, F., {Leroy}, A., {Walter}, F., {Brinks}, E., {de Blok}, W.~J.~G.,
  {Madore}, B., \& {Thornley}, M.~D. 2008, \aj, 136, 2846

\bibitem[{{Bigiel} {et~al.}(2011){Bigiel}, {Leroy}, {Walter}, {Brinks}, {de
  Blok}, {Kramer}, {Rix}, {Schruba}, {Schuster}, {Usero}, \&
  {Wiesemeyer}}]{sfr:blwb11}
{Bigiel}, F., {Leroy}, A.~K., {Walter}, F., {Brinks}, E., {de Blok}, W.~J.~G.,
  {Kramer}, C., {Rix}, H.~W., {Schruba}, A., {Schuster}, K., {Usero}, A., \&
  {Wiesemeyer}, H.~W. 2011, \apjl, 730, L13+

\bibitem[{{Bolatto} {et~al.}(2011){Bolatto}, {Leroy}, {Jameson}, {Ostriker},
  {Gordon}, {Lawton}, {Stanimirovic}, {Israel}, {Madden}, {Hony}, {Sandstrom},
  {Bot}, {Rubio}, {Winkler}, {Roman-Duval}, {van Loon}, {Oliveira}, \&
  {Indebetouw}}]{sfr:bljo11}
{Bolatto}, A.~D., {Leroy}, A.~K., {Jameson}, K., {Ostriker}, E., {Gordon}, K.,
  {Lawton}, B., {Stanimirovic}, S., {Israel}, F.~P., {Madden}, S.~C., {Hony},
  S., {Sandstrom}, K.~M., {Bot}, C., {Rubio}, M., {Winkler}, P.~F.,
  {Roman-Duval}, J., {van Loon}, J.~T., {Oliveira}, J.~M., \& {Indebetouw}, R.
  2011, ArXiv e-prints

\bibitem[{{Daddi} {et~al.}(2010){Daddi}, {Bournaud}, {Walter}, {Dannerbauer},
  {Carilli}, {Dickinson}, {Elbaz}, {Morrison}, {Riechers}, {Onodera}, {Salmi},
  {Krips}, \& {Stern}}]{sfr:dbwd10}
{Daddi}, E., {Bournaud}, F., {Walter}, F., {Dannerbauer}, H., {Carilli}, C.~L.,
  {Dickinson}, M., {Elbaz}, D., {Morrison}, G.~E., {Riechers}, D., {Onodera},
  M., {Salmi}, F., {Krips}, M., \& {Stern}, D. 2010, \apj, 713, 686

\bibitem[{{Feldmann} {et~al.}(2011){Feldmann}, {Gnedin}, \&
  {Kravtsov}}]{ng:fgk11}
{Feldmann}, R., {Gnedin}, N.~Y., \& {Kravtsov}, A.~V. 2011, \apj, 732, 115

\bibitem[{{Feldmann} {et~al.}(2012){Feldmann}, {Gnedin}, \&
  {Kravtsov}}]{ng:fgk12b}
---. 2012, \apj, 758, 127

\bibitem[{{Genzel} {et~al.}(2010){Genzel}, {Tacconi}, {Gracia-Carpio},
  {Sternberg}, {Cooper}, {Shapiro}, {Bolatto}, {Bouch{\'e}}, {Bournaud},
  {Burkert}, {Combes}, {Comerford}, {Cox}, {Davis}, {Schreiber},
  {Garcia-Burillo}, {Lutz}, {Naab}, {Neri}, {Omont}, {Shapley}, \&
  {Weiner}}]{sfr:gtgs10}
{Genzel}, R., {Tacconi}, L.~J., {Gracia-Carpio}, J., {Sternberg}, A., {Cooper},
  M.~C., {Shapiro}, K., {Bolatto}, A., {Bouch{\'e}}, N., {Bournaud}, F.,
  {Burkert}, A., {Combes}, F., {Comerford}, J., {Cox}, P., {Davis}, M.,
  {Schreiber}, N.~M.~F., {Garcia-Burillo}, S., {Lutz}, D., {Naab}, T., {Neri},
  R., {Omont}, A., {Shapley}, A., \& {Weiner}, B. 2010, \mnras, 407, 2091

\bibitem[{{Hopkins} {et~al.}(2013){Hopkins}, {Keres}, {Onorbe},
  {Faucher-Giguere}, {Quataert}, {Murray}, \& {Bullock}}]{sims:hko13}
{Hopkins}, P.~F., {Keres}, D., {Onorbe}, J., {Faucher-Giguere}, C.-A.,
  {Quataert}, E., {Murray}, N., \& {Bullock}, J.~S. 2013, ArXiv e-prints

\bibitem[{{Hopkins} {et~al.}(2011){Hopkins}, {Quataert}, \&
  {Murray}}]{sims:hqm11}
{Hopkins}, P.~F., {Quataert}, E., \& {Murray}, N. 2011, \mnras, 417, 950

\bibitem[{{Kravtsov}(2003)}]{misc:k03}
{Kravtsov}, A.~V. 2003, \apjl, 590, L1

\bibitem[{{Kritsuk} {et~al.}(2013){Kritsuk}, {Lee}, \& {Norman}}]{ism:kln13}
{Kritsuk}, A.~G., {Lee}, C.~T., \& {Norman}, M.~L. 2013, \mnras, 436, 3247

\bibitem[{{Kritsuk} {et~al.}(2011){Kritsuk}, {Norman}, \& {Wagner}}]{ism:knw11}
{Kritsuk}, A.~G., {Norman}, M.~L., \& {Wagner}, R. 2011, \apjl, 727, L20

\bibitem[{{Krumholz} \& {Tan}(2007)}]{sfr:kt07}
{Krumholz}, M.~R. \& {Tan}, J.~C. 2007, \apj, 654, 304

\bibitem[{{Lada} {et~al.}(2010){Lada}, {Lombardi}, \& {Alves}}]{sfr:lla10}
{Lada}, C.~J., {Lombardi}, M., \& {Alves}, J.~F. 2010, \apj, 724, 687

\bibitem[{{Leroy} {et~al.}(2012){Leroy}, {Bigiel}, {de Blok}, {Boissier},
  {Bolatto}, {Brinks}, {Madore}, {Munoz-Mateos}, {Murphy}, {Sandstrom},
  {Schruba}, \& {Walter}}]{sfr:lbbb12}
{Leroy}, A.~K., {Bigiel}, F., {de Blok}, W.~J.~G., {Boissier}, S., {Bolatto},
  A., {Brinks}, E., {Madore}, B., {Munoz-Mateos}, J.-C., {Murphy}, E.,
  {Sandstrom}, K., {Schruba}, A., \& {Walter}, F. 2012, \aj, 144, 3

\bibitem[{{Leroy} {et~al.}(2008){Leroy}, {Walter}, {Brinks}, {Bigiel}, {de
  Blok}, {Madore}, \& {Thornley}}]{ism:lwbb08}
{Leroy}, A.~K., {Walter}, F., {Brinks}, E., {Bigiel}, F., {de Blok}, W.~J.~G.,
  {Madore}, B., \& {Thornley}, M.~D. 2008, \aj, 136, 2782

\bibitem[{{Leroy} {et~al.}(2013){Leroy}, {Walter}, {Sandstrom}, {Schruba},
  {Munoz-Mateos}, {Bigiel}, {Bolatto}, {Brinks}, {de Blok}, {Meidt}, {Rix},
  {Rosolowsky}, {Schinnerer}, {Schuster}, \& {Usero}}]{sfr:lwss13}
{Leroy}, A.~K., {Walter}, F., {Sandstrom}, K., {Schruba}, A., {Munoz-Mateos},
  J.-C., {Bigiel}, F., {Bolatto}, A., {Brinks}, E., {de Blok}, W.~J.~G.,
  {Meidt}, S., {Rix}, H.-W., {Rosolowsky}, E., {Schinnerer}, E., {Schuster},
  K.-F., \& {Usero}, A. 2013, ArXiv e-prints

\bibitem[{{McKee} \& {Ostriker}(2007)}]{sfr:mo07}
{McKee}, C.~F. \& {Ostriker}, E.~C. 2007, \araa, 45, 565

\bibitem[{{Padoan} \& {Nordlund}(2011)}]{ism:pn11}
{Padoan}, P. \& {Nordlund}, {\AA}. 2011, \apj, 730, 40

\bibitem[{{Schaye} {et~al.}(2009){Schaye}, {Vecchia}, {Booth}, {Wiersma},
  {Theuns}, {Haas}, {Bertone}, {Duffy}, {McCarthy}, \& {van de
  Voort}}]{sims:svbw09}
{Schaye}, J., {Vecchia}, C.~D., {Booth}, C.~M., {Wiersma}, R.~P.~C., {Theuns},
  T., {Haas}, M.~R., {Bertone}, S., {Duffy}, A.~R., {McCarthy}, I.~G., \& {van
  de Voort}, F. 2009, \mnras, 1888

\bibitem[{{Schruba} {et~al.}(2010){Schruba}, {Leroy}, {Walter}, {Sandstrom}, \&
  {Rosolowsky}}]{sfr:slw10}
{Schruba}, A., {Leroy}, A.~K., {Walter}, F., {Sandstrom}, K., \& {Rosolowsky},
  E. 2010, \apj, 722, 1699

\bibitem[{{Tacconi} {et~al.}(2013){Tacconi}, {Neri}, {Genzel}, {Combes},
  {Bolatto}, {Cooper}, {Wuyts}, {Bournaud}, {Burkert}, {Comerford}, {Cox},
  {Davis}, {F{\"o}rster Schreiber}, {Garc{\'{\i}}a-Burillo}, {Gracia-Carpio},
  {Lutz}, {Naab}, {Newman}, {Omont}, {Saintonge}, {Shapiro Griffin}, {Shapley},
  {Sternberg}, \& {Weiner}}]{sfr:tngc13}
{Tacconi}, L.~J., {Neri}, R., {Genzel}, R., {Combes}, F., {Bolatto}, A.,
  {Cooper}, M.~C., {Wuyts}, S., {Bournaud}, F., {Burkert}, A., {Comerford}, J.,
  {Cox}, P., {Davis}, M., {F{\"o}rster Schreiber}, N.~M.,
  {Garc{\'{\i}}a-Burillo}, S., {Gracia-Carpio}, J., {Lutz}, D., {Naab}, T.,
  {Newman}, S., {Omont}, A., {Saintonge}, A., {Shapiro Griffin}, K., {Shapley},
  A., {Sternberg}, A., \& {Weiner}, B. 2013, \apj, 768, 74

\bibitem[{{Tasker}(2011)}]{Tasker2011}
{Tasker}, E.~J. 2011, \apj, 730, 11

\bibitem[{{The Enzo Collaboration} {et~al.}(2013){The Enzo Collaboration},
  {Bryan}, {Norman}, {O'Shea}, {Abel}, {Wise}, {Turk}, {Reynolds}, {Collins},
  {Wang}, {Skillman}, {Smith}, {Harkness}, {Bordner}, {Kim}, {Kuhlen}, {Xu},
  {Goldbaum}, {Hummels}, {Kritsuk}, {Tasker}, {Skory}, {Simpson}, {Hahn},
  {Oishi}, {So}, {Zhao}, {Cen}, \& {Li}}]{Enzo2013}
{The Enzo Collaboration}, {Bryan}, G.~L., {Norman}, M.~L., {O'Shea}, B.~W.,
  {Abel}, T., {Wise}, J.~H., {Turk}, M.~J., {Reynolds}, D.~R., {Collins},
  D.~C., {Wang}, P., {Skillman}, S.~W., {Smith}, B., {Harkness}, R.~P.,
  {Bordner}, J., {Kim}, J.-h., {Kuhlen}, M., {Xu}, H., {Goldbaum}, N.,
  {Hummels}, C., {Kritsuk}, A.~G., {Tasker}, E., {Skory}, S., {Simpson}, C.~M.,
  {Hahn}, O., {Oishi}, J.~S., {So}, G.~C., {Zhao}, F., {Cen}, R., \& {Li}, Y.
  2013, ArXiv e-prints

\end{thebibliography}

\end{document}